\documentclass[aps,prl,reprint,superscriptaddress]{revtex4-1}
\usepackage{graphics}
\begin{document}

\title{Quenched Slonczewski-Windmill in Spin-Torque Vortex-Oscillators}

\author{V. Sluka}
\altaffiliation[present adress: ]{Institute of Ion Beam Physics and Materials Research, Helmholtz-Zentrum Dresden-Rossendorf e. V., D-01314 Dresden, Germany}
\affiliation{Peter Gr\"unberg Institute, Electronic Properties (PGI-6) and J\"ulich-Aachen Research Alliance, Fundamentals of Future Information Technology (JARA-FIT), Forschungszentrum J\"ulich GmbH, D-52425 J\"ulich, Germany}
\author{A. K\'{a}kay}
\affiliation{Peter Gr\"unberg Institute, Electronic Properties (PGI-6) and J\"ulich-Aachen Research Alliance, Fundamentals of Future Information Technology (JARA-FIT), Forschungszentrum J\"ulich GmbH, D-52425 J\"ulich, Germany}
\author{A. M. Deac}
\affiliation{Institute of Ion Beam Physics and Materials Research, Helmholtz-Zentrum Dresden-Rossendorf e. V., D-01314 Dresden, Germany}
\author{D. E. B\"urgler}
\affiliation{Peter Gr\"unberg Institute, Electronic Properties (PGI-6) and J\"ulich-Aachen Research Alliance, Fundamentals of Future Information Technology (JARA-FIT), Forschungszentrum J\"ulich GmbH, D-52425 J\"ulich, Germany}
\author{R. Hertel}
\affiliation{Institut de Physique et Chimie des Mat\'{e}riaux de Strasbourg, Universit\'{e} de Strasbourg, CNRS UMR 7504, F-67034 Strasbourg Cedex 2, France}
\author{C. M. Schneider}
\affiliation{Peter Gr\"unberg Institute, Electronic Properties (PGI-6) and J\"ulich-Aachen Research Alliance, Fundamentals of Future Information Technology (JARA-FIT), Forschungszentrum J\"ulich GmbH, D-52425 J\"ulich, Germany}

\date{\today}

\begin{abstract}
We present a combined analytical and numerical study on double-vortex
spin-torque nano-oscillators and describe a mechanism that suppresses
the windmill modes. The magnetization dynamics is dominated by the gyrotropic precession of the vortex in one of the ferromagnetic layers.  In the other layer the vortex gyration is strongly damped.  The dominating layer for the magnetization dynamics is determined by the current polarity.
Measurements on Fe/Ag/Fe nano-pillars support these findings. The results open up a new perspective for building high quality-factor spin-torque oscillators operating at selectable, well-separated frequency bands.
\end{abstract}

\pacs{}

\maketitle

The advent of spintronics lead to the development of exciting new concepts for nano-scale devices using the spin-degree of freedom of the electron besides its charge-property \cite{Wol2001}. Much effort has been spent on spin-torque nano-oscillators (STNOs) \cite{Kis2003,Kri2005,Man2005,Kak2005,Sil2007,Dea2008}, which typically consist of two single domain ferromagnetic layers separated by a metallic spacer or a tunnel barrier, one with its magnetization fixed (polarizing layer), the other one susceptible to torques (free layer). An electric current traversing the system perpendicular to the layers becomes spin-polarized and exerts torques on the magnetizations \cite{Slo1996,Ber1996,Slo2002}, leading to magnetization dynamics of the free layer.  These excitations are typically in the range of a few gigahertz and can be
detected by measuring the time variation of the magnetoresistance
(MR).  The pinning of the polarizing layer can, for example, be
achieved by exchange coupling to an antiferromagnet \cite{Kri2005} or
by extending its thickness and lateral dimension \cite{Kis2003}. In
the absence of pinning both ferromagnetic layers will be excited and
in the case of increasingly symmetric STNOs, this results in a dynamic
equilibrium state called the Slonczewski-windmill
\cite{Slo1996,Baz2008}. In this state both layers' magnetizations
rotate in the same direction with a constant relative angle, resulting
in a vanishing MR time-dependence.\\
 Here we investigate STNOs containing two stacked magnetic vortices, i.e., a system consisting of two ferromagnetic disks, each in a vortex state and separated by a metallic, nonmagnetic spacer. Employing analytical and numerical methods, we study the coupled spin torque-driven motion of the magnetizations in the two disks, which are not pinned by any of the above mentioned mechanisms. We find that in the double vortex system, Slonczewski-windmill modes are quenched by an intriguing mechanism. Our results show that that the current polarity determines which disk is excited and thereby selects the
STNO-frequency band. This property is shown to arise from a spin torque-mediated vortex-vortex interaction. Thus, it is an entirely different principle than the spin accumulation
based mechanism suggested by Tsoi et al.\,\cite{Tso2004}. We analyze in detail the underlying torques and the resulting forces. The force exerted by one vortex onto the other can be split into two contributions; one part arising from the polarizer vortex in-plane magnetization acting on the free vortex core (disk-core part) and another one which is due to the core-disk interaction. We compute the dependence of these terms on the lateral core-core distance. These results provide insight into the fascinating dynamics of coupled magnetic vortices. The theoretical findings are supported by our experimental data obtained from double-vortex Fe/Ag/Fe STNOs.\\
 The motion of the magnetic vortex in each of the disks is governed by the Thiele equation \cite{Thi1973} which we write here for the vortex in the top disk,
\begin{equation}
\mathbf{G}_{1}\times \frac{\mathrm{d}\mathbf{X}_{1}}{\mathrm{d}t}-\frac{\mathrm{d}W_{1}}{\mathrm{d}\mathbf{X}_{1}}-D_{1}\frac{\mathrm{d}\mathbf{X}_{1}}{\mathrm{d}t}+\frac{\hbar jP}{4e}\mathbf{F}_{1}=0.
\end{equation}
$\mathbf{G}=-2\pi\kappa(\mu_{0}M_{s}L/\gamma)\mathbf{\hat{e}}_{z}$ is the gyro vector, where $L$ and $\kappa$ are the disk thickness and the vortex core polarity, respectively. $\mathbf{X}$ is the core position with respect to the disk center, $W$ refers to the effective magnetostatic potential in which the core is moving, and $D=(\alpha\mu_{0}M_{s}/\gamma)L\pi\ln (R/r_{0})$ characterizes the damping of the vortex motion. The parameters $R$ and $r_{0}$ are the radii of the disk and the vortex core, and the indices 1 and 2 correspond to the top and bottom disks, respectively. The spin-transfer torque-induced force acting on the vortex generated by a vortex-polarizer can be decomposed into two contributions $\mathbf{F}_{1}=\mathbf{F}_{1}^{d}+\mathbf{F}_{1}^{c}$. $\mathbf{F}_{1}^{d}$ arises from the in-plane magnetization of the polarizer and acts on the core of the free vortex. The second term $\mathbf{F}_{1}^{c}$ is caused by the core of the polarizer-vortex and acts on the in-plane magnetization of the free vortex. Both force contributions depend on the lateral core-core distance $l$.
Following Ref. \cite{Iva2007}, we obtain the expressions
\begin{figure}[h!]
\includegraphics{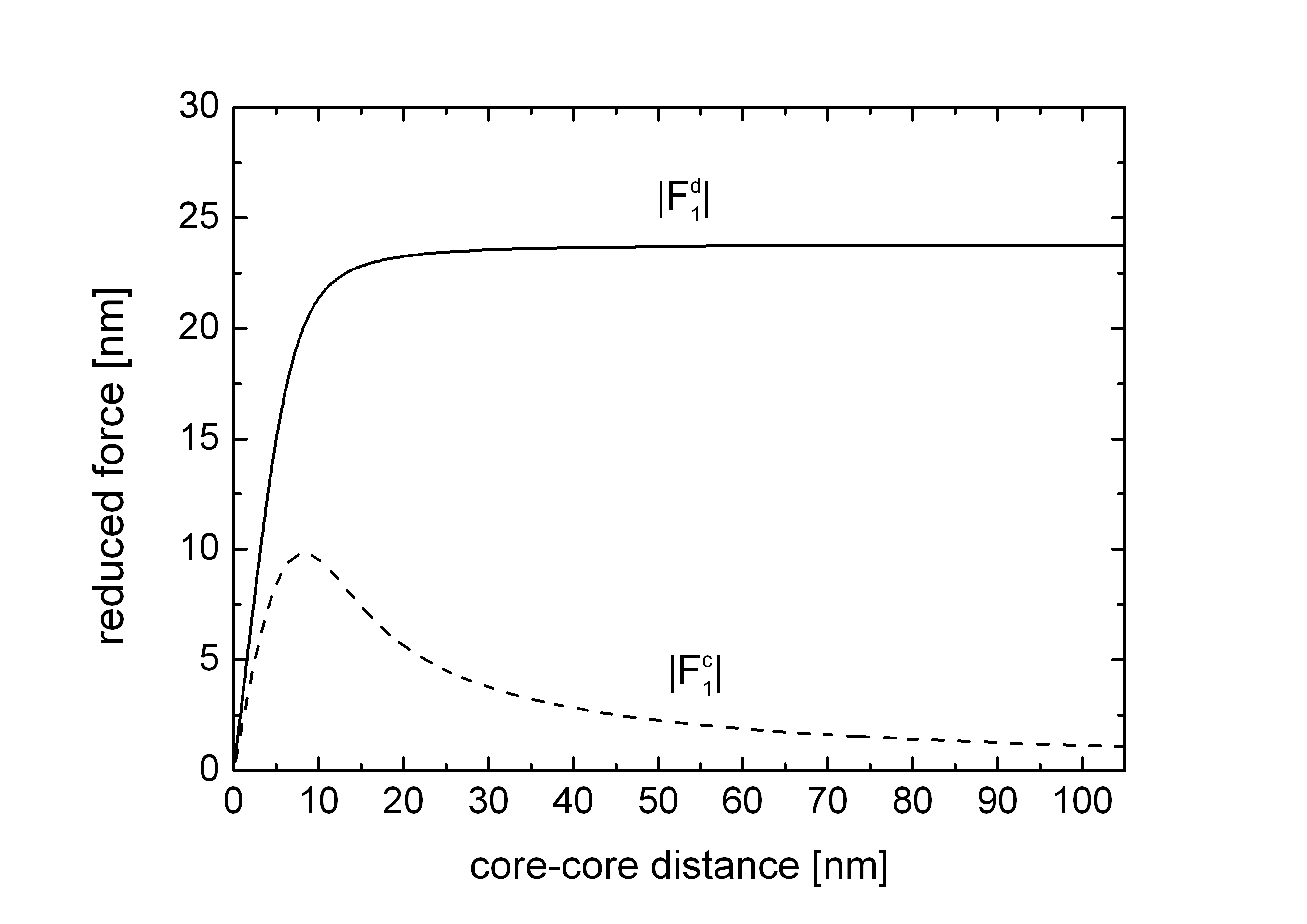}
\caption{\label{fig-1} Spin-torque magnitudes exerted by disk 2 on the vortex in disk 1 and their dependence on the lateral core-core separation. The torque arising from the polarizer core, $F^{c}_{1}$ is negligible at large distances, explaining the results obtained by micromagnetic simulations and presented in Ref. \cite{Khv2010}.}
\end{figure}
\begin{eqnarray}
\mathbf{F}_{1}^{d}&=& \int_{A}\mathrm{d}^{2}x\,\left[m_{2,x}\sin(\varphi)-m_{2,y}\cos(\varphi)\right]\nabla \theta\nonumber \\
& +&\sin(\varphi)\cos(\varphi)\left[m_{2,x}\cos(\varphi)+m_{2,y}\sin(\varphi)\right]\nabla \varphi\nonumber\\
&=&C_{1}C_{2}\kappa_{1}F^{d}\mathbf{\hat{e}}_{12},
\end{eqnarray}
and
\begin{equation}
\mathbf{F}_{1}^{c}=-\int_{A}\mathrm{d}^{2}x\, m_{2,z}\sin^{2}(\theta)\nabla \varphi=\kappa_{2}F^{c}\mathbf{\hat{e}}_{12}\label{fc},
\end{equation}
where we introduce the vectors
\begin{equation}
\mathbf{\hat{e}}_{12}=-\mathbf{\hat{e}}_{21}=\frac{\mathbf{\hat{e}}_{z}\times(\mathbf{X}_{1}-
\mathbf{X}_{2})}{\sqrt{(X_{1}-X_{2})^{2}+(Y_{1}-Y_{2})^{2}}}.
\end{equation}
$C_{i}=\pm 1$ defines the vorticity (``+" corresponds to counterclockwise, ``-" to clockwise) and $m_{2,i}$ refers to the $i$th component of the bottom vortex unit magnetization vector. The top layer magnetization is written in spherical coordinates $(\varphi,\theta)$, where $\varphi$ and $\theta$ are the azimuthal and polar angles, respectively. In the calculation we assume rigid vortices with thickness-independent magnetization. The cylinder axis is chosen as the $z$-axis.
Using the ansatz of Feldtkeller and Thomas \cite{Fel1965} for the out-of-plane core magnetization ($|m_{i,z}|=\exp{(-a^{2}r^{2})}$ with $a^{2}\approx \ln 2/(\mathrm{25\,nm^{2}})$ to mimic the  experimentally obtained vortex core size in Fe \cite{Wac2002}), we obtain the forces and their dependence on the lateral core-core distance $l$ as shown in Fig. \ref{fig-1}. For large $l$, the contribution of $\mathbf{F}_{1}^{c}$ to the total force becomes negligible. This explains the simulation results reported in Ref. \cite{Khv2010}, where the influence of the polarizer-vortex core on the dynamics was found to be small. From Eq. (\ref{fc}) we see that this is caused by the reduction of the function $|\nabla \varphi|$ with increasing distance from the top vortex core. In contrast to the asymptotic decrease of $\mathbf{F}^{c}_{1}$ to zero, the magnitude of $\mathbf{F}^{d}_{1}$ approaches a finite value for large $l$. For small distances $l<2r_{0}\approx \mathrm{10\,nm}$, however, we observe that both torques fall to zero. This can be attributed to the gain in symmetry with decreasing core-core distance. The small torque introduced by  $\mathbf{F}^{c}_{1}$ is neglected in our investigation on the dynamics of the system with the vortices coupled by the electric current. This is justified by the fact that we are interested in the general behavior of the solutions. The decrease of $\mathbf{F}_{1}^{d}$ at small $l$ must however be included. The coupled Thiele equations read, with $\tilde{j}:=\hbar j P /(4e)$
\begin{eqnarray}
\mathbf{G}_{1}\times \frac{\mathrm{d}\mathbf{X}_{1}}{\mathrm{d}t}-\frac{\mathrm{d}W_{1}}{\mathrm{d}\mathbf{X}_{1}}
-D_{1}\frac{\mathrm{d}\mathbf{X}_{1}}{\mathrm{d}t}+\tilde{j}C_{1}C_{2}\kappa_{1}F^{d}
\mathbf{\hat{e}}_{12}=0\,\,\,\,\,\,\,\,\\
\mathbf{G}_{2}\times \frac{\mathrm{d}\mathbf{X}_{2}}{\mathrm{d}t}-\frac{\mathrm{d}W_{2}}{\mathrm{d}\mathbf{X}_{2}}
-D_{2}\frac{\mathrm{d}\mathbf{X}_{2}}{\mathrm{d}t}-\tilde{j}C_{1}C_{2}\kappa_{2}F^{d}
\mathbf{\hat{e}}_{21}=0.\,\,\,\,\,\,\,\,
\end{eqnarray}
The sign of $\tilde{j}$ is positive for electron
flow from the top to the bottom layer.
For a quantitative analysis of the solutions, we use parabolic approximations to the effective magnetostatic potentials. We let $W_{1}(\mathbf{X}_{1})/|\mathbf{G}_{1}|=6.28\,\mathrm{ns}^{-1}\mathbf{X}_{1}^{2}/2$, resulting in a top vortex eigenfrequency of $f^{0}_{1}=\mathrm{1.0\,GHz}$.
The bottom disk potential is fixed at $W_{2}/|\mathbf{G}_{2}|=(5/3)W_{1}/|\mathbf{G}_{1}|$ ($f^{0}_{2}=\mathrm{1.7\,GHz}$). These frequencies 
are chosen to represent our Fe/Ag/Fe nanopillars with ferromagnetic 
layers with a thickness ratio of 5/3.
We use a current of $\tilde{j}/|\mathbf{G}_{1}|=\mathrm{(5/23)\,ns^{-1}}$ corresponding to about $\mathrm{1.12\times 10^{12}A/m^{2}}$ (for $P=1$), which is within the range of experiments. The spin-transfer torque-induced force $F^{d}$ is assumed to increase linearly from $l=0$ to $l=\mathrm{10\,nm}$, from whereon it is set to the constant value $F^{d}_{\infty}=\mathrm{23\,nm}$.\\
The solutions are obtained numerically using Maple's rkf45 implementation \cite{Map2011}. 
The results can be summarized as follows:  For
positive  currents  and  equal  vorticities, the top vortex gyrates around the disk center on a trajectory of about $\mathrm{50\,nm}$ in radius, regardless  of  the  core polarity. The sense of rotation
is  determined by the  core polarization (counterclockwise for
positive  and clockwise  for  negative core  polarity). The  gyration
frequency is $\mathrm{1.0\,GHz}$. The
bottom vortex  adapts its frequency and  sense of gyration according to the top
vortex.   A dynamic  equilibrium develops with  a  constant phase
difference between the vortices even  in the case of opposite relative
core  orientation. This frequency  adaption is accompanied by  a strong
reduction in radius  of the  bottom core  trajectory:  for parallel
cores,  the   radius  is   about  $\mathrm{0.7\,nm}$  while   for  the
antiparallel  configuration, the reduction is even  more pronounced
(approximately  $\mathrm{0.2\,nm}$).  For  negative  currents, the  vortices
switch   roles:  The   bottom  vortex   gyrates  on   a   large  orbit
($\mathrm{18\,nm}$),  while  the top  vortex  trajectory is  quenched
($\mathrm{1.2\,nm}$ for  parallel, $\mathrm{0.3\,nm}$ for antiparallel
core alignment).   In the dynamic equilibrium the  phase difference is
constant and the gyration frequency is about $\mathrm{1.7\,GHz}$. This
frequency corresponds to the eigenfrequency of the bottom vortex while
the sense of rotation is determined by its core polarity.
\begin{figure}[h!]
\includegraphics{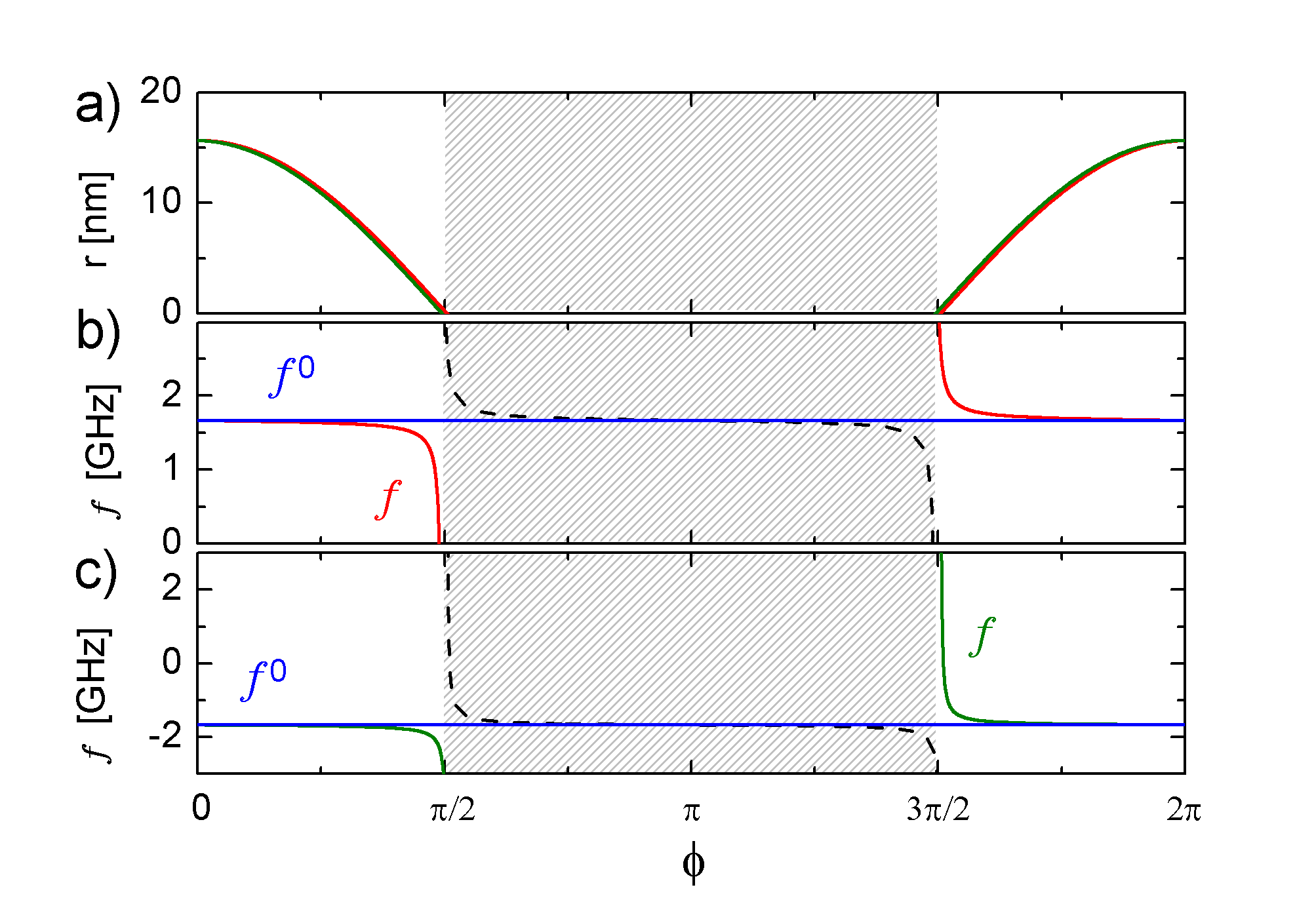}
\caption{\label{fig-2} Trajectory  radius $r$ and frequency  $f$ of the
bottom vortex as functions of the phase $\phi$ with 
respect to the top vortex that
gyrates at the same frequency 
for positive (red) and negative (green)
core  polarity. The blue lines mark  the eigenfrequency $f^{0}$ of
the  free running bottom vortex for positive  (b) and  negative (c) core
polarity. The shaded regions correspond to equivalent solutions but
negative radius.}
\end{figure}
The gyration phase difference between the
two  vortices  depends  on  the  relative core  alignment. For
positive  currents  and  parallel   cores,  the  bottom  core  gyrates
approximately $\mathrm{90^\circ}$ ahead  of its top counterpart, while
for antiparallel  cores a  $\mathrm{90^\circ}$ lag is  observed. 
From Eqs. (5) and (6)  it is  clear that  the solutions  for opposite
vorticities are identical to those obtained for equal vorticities with
a negative current polarity.\\
 For large enough $|j|$, the obtained characteristics
of the dynamics are the generalization of the criterion found in
Ref.\,\cite{Khv2010}.  In the model used by those authors, the polarizer was assumed to be
a fixed, rigid vortex, and only magnetization dynamics in the other, free
disk was allowed.  In our case, both disks can be polarizing or free
layer.  For a given combination of vorticities $C_{1}C_{2}$ and
applied current polarity, the system responds with a damped and a
dominant gyration, the former defining the polarizing and the latter
the free disk.  The current polarity determines which disk is
dominantly excited.  Therefore, the generalized $jCC$-criterion reads:
For $jC_{1}C_{2}>0$ the top and for $jC_{1}C_{2}<0$ the bottom disk is
excited.\\
 A general analytical solution for steady state trajectories is difficult to establish due to the complexity of the denominator in Eq. (4); however, it is instructive to analyze the situation when one trajectory radius is much larger than the other, a condition that, according to our numerical results, also holds in the vicinity of the limit cycle. We consider the case of positive $jC_{1}C_{2}$ and replace the denominator in Eq. (4) by $d:=\sqrt{X_{1}^{2}+Y_{1}^{2}}$, yielding the following relations between the radius $r$ of the bottom vortex trajectory, its phase $\phi$ (relative to the top vortex) and the common frequency $f$ of the two oscillators:
\begin{figure}[h!]
\includegraphics{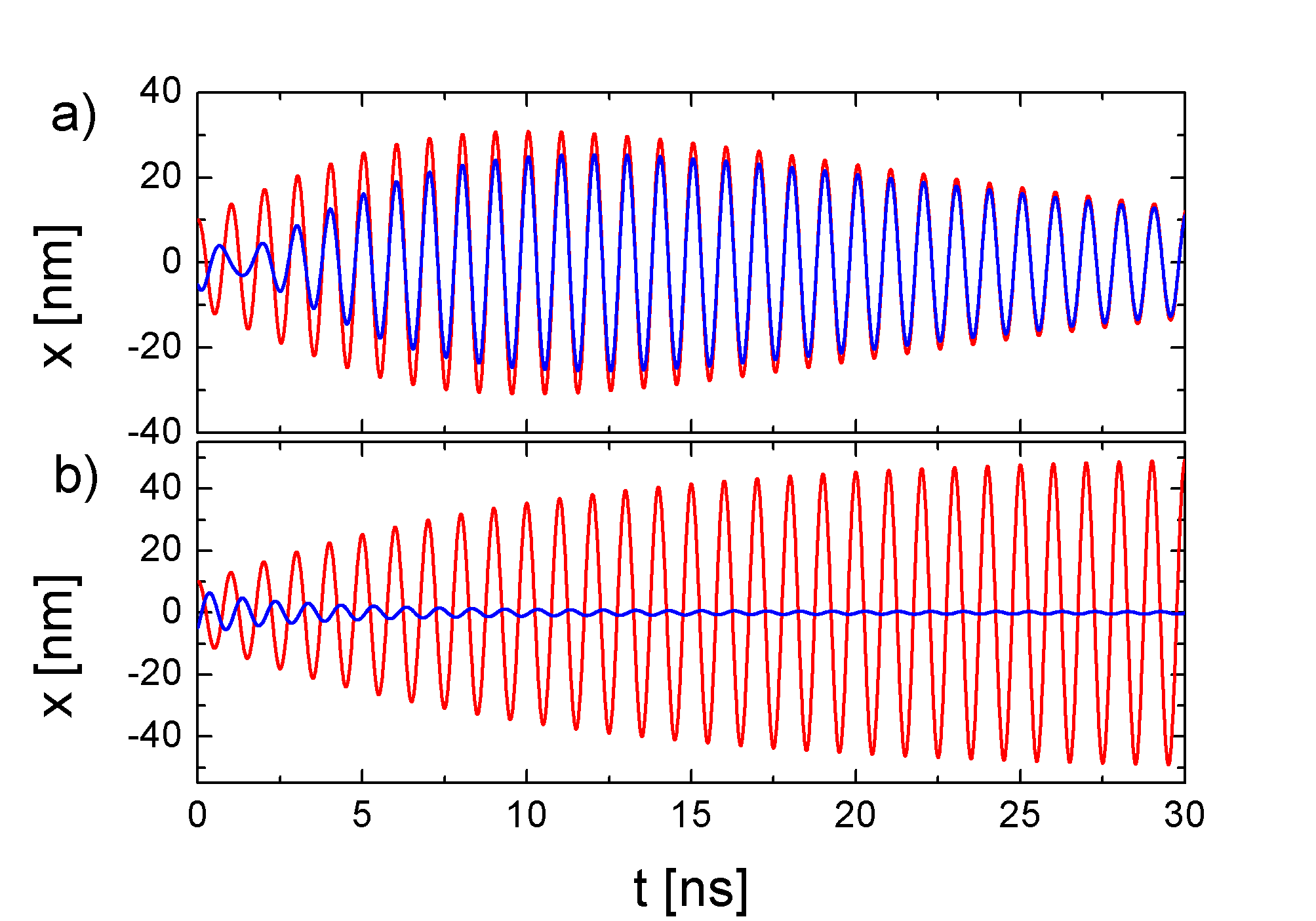}
\caption{\label{fig-3} $X$-components of the top (red) and bottom
(blue) core coordinates versus time for the case of symmetric disks.
The cores are aligned parallel in (a) and antiparallel in (b).  In the
parallel case, a windmill-mode only appears in a transient time
interval, but is hindered afterwards.  The oscillation decays due to the low core-core 
separation and the related decrease of the spin-transfer
torque-induced force.}
\end{figure}
\begin{equation}
r(\phi)=\frac{\frac{\tilde{j}C_{1}C_{2}F^{d}_{\infty}}{2\pi D_{2}}\cos\phi+\frac{\kappa_{2}\tilde{j}C_{1}C_{2}F^{d}_{\infty}}{2\pi |\mathbf{G}_{2}|}\sin\phi}{f_{2}^{0}+\frac{\tilde{j}C_{1}C_{2}F^{d}_{\infty}}{2\pi D_{2}d}}
\end{equation}
\begin{equation}
f(\phi)=f_{2}^{0}-\frac{\tilde{j}C_{1}C_{2}F^{d}_{\infty}}{2\pi|\mathbf{G}_{2}|
r(\phi)}\sin\phi
\end{equation}
These relations are displayed in Fig. \ref{fig-2} and reproduce the behavior observed in the numerical solutions: For both cases of positive and negative bottom vortex core polarity, the bottom vortex can adapt to the ({\it a priori} arbitrary) frequency of the top vortex by adjusting the phase. Positive (negative) frequency corresponds to counterclockwise (clockwise) gyration. As displayed in Fig. \ref{fig-2}(a), this phase shifting comes with a strong reduction of the orbit radius- or, in other words, a quenching of the windmill-modes.
The dashed lines in the shaded regimes correspond to solutions of negative radius. Since a reverse of the sign of the radius is equivalent to a phase shift of $\pi$, these negative-$r$ solutions are identical to the trajectories represented by the solid lines. By means of a phase adaption and reduction of the radius, the vortex can use a fraction of the spin-transfer torque-induced force to assist or counteract the force due to its magnetostatic potential. The resulting radial force component can differ strongly from the purely magnetostatic force. It may even lead to an inversion of the relation between the sense of gyration and the core polarity.
A special case is the configuration, for which a frequency adaption is not necessary, i.e., if the two disks are identical and the cores parallel [Fig. \ref{fig-3}a)].
\begin{figure}[h!]
\includegraphics{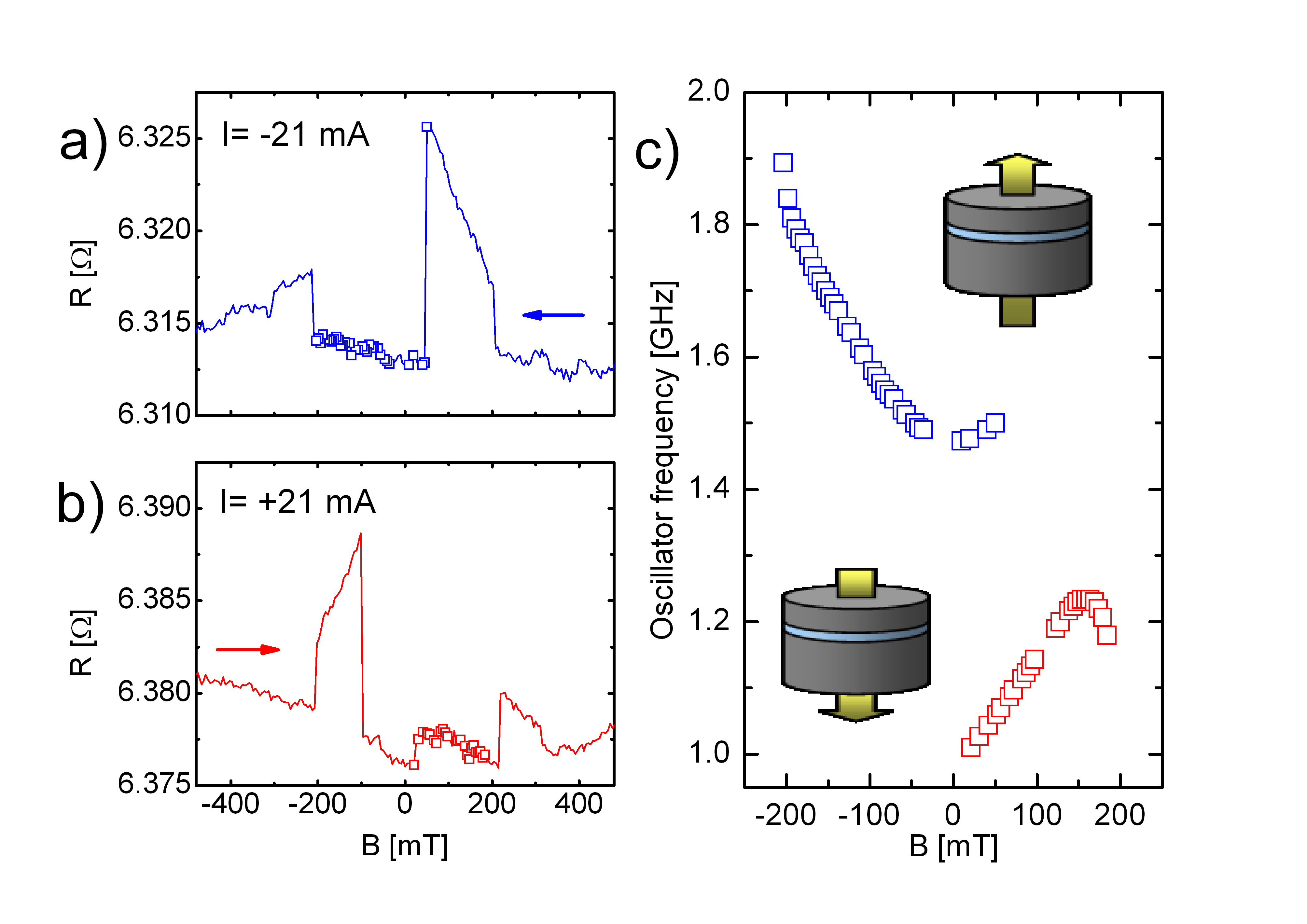}
\caption{\label{fig-4} Resistance versus field for
(a) $I=-21$ and (b) $I=+21$\,mA sample current. $I<0$ corresponds to electron flow from bottom to top
electrode. Squares on the curves indicate the observation of
magnetization dynamics for the corresponding field value.  (c) Field
dependence of the double-vortex mode frequency for the two electron
flow directions as indicated.}
\end{figure}
In this case, the vortices start rotating in phase, but as they reach the limit cycle and the core-core distance drops below $\mathrm{10\,nm}$, the mutual spin-transfer torque-induced force decreases leading to a decay of the oscillation amplitudes.
For antiparallel alignment, the windmill-modes are quenched by the mechanism of frequency and phase adaption [Fig. \ref{fig-3}b].\\
 For an experimental confirmation of the frequency
and phase adaption mechanism and the related quenching of the windmill
modes we study the current-induced magnetization dynamics of an
Fe/Ag/Fe nanopillar with a Fe layer thickness ratio of 5/3.  According
to our model, we expect to observe excitations for both current
polarities, but with different frequencies yielding a frequency ratio
of approximately 5/3.  Cylindrical nanopillars are patterned using
e-beam lithography and Ar ion milling from molecular beam
epitaxy-grown GaAs(001)/ Fe(1)/Ag(150)/Fe(25)/Ag(6)/Fe(15)/Au(25)
stacks (layer thicknesses in nm).  The pillar diameter is
$\mathrm{210\,nm}$.  The milling was stopped after reaching the
$\mathrm{150\,nm}$ thick Ag buffer layer.  Thus, the oscillator
consists of two ferromagnetic disks of equal diameter and comparable
thickness stacked on top of each other.  Figures \ref{fig-4}(a) and (b)
display the field dependence of the nanopillar resistance for
$I=\mp 21$\,mA ($\mp 6.1\times 10^{7}$\,A/cm$^{2}$), respectively.  The
external magnetic field was applied in the sample plane.  The
magnetoresistance profiles are characteristic for this sample type
\cite{Slu2011} 
and reflect two magnetization states: The first
one comprises a vortex in one disk, while the other nanomagnet remains
in a quasi-homogeneous state.  These configurations are characterized
by a nearly linear field dependence of the resistance caused by a
continuous lateral displacement of the vortex with changing field.
The second state is observed from low field magnitudes up to about
$\mathrm{200\,mT}$ and is characterized by low resistance values near
the level in magnetic saturation.  Here, each disk contains a vortex with
the vorticity given by the circumferential Oersted field.  This
results in locally parallel alignment of the two disks' magnetizations
explaining the observed low resistance.  For both current polarities
in Figs.\,\ref{fig-4}(a) and (b) we detected magnetization dynamics in
those field intervals, in which the double-vortex state occurs.  The
excitation frequencies are shown in Fig. \ref{fig-4}(c) along with the
corresponding electron flow directions of the externally applied
currents $I$.  All frequencies are below $\mathrm{2\,GHz}$, which is
typical for vortex gyration in Fe/Ag/Fe nanopillars
\cite{Leh2009,Leh2010,Slu2011}, but the
frequencies are clearly different and well separated for the two
current polarities.  At low external fields their ratio is about
$\mathrm{1.46}$. Using previous data on single vortex dynamics
\cite{Khv2009,Slu2011}, we estimate the influence of the Oersted field
on the zero-field vortex frequencies to be $\simeq\mathrm{145}$ MHz for both disks. This shifts the above value to $\mathrm{1.54}$, {\it i.e.} very close to the ratio of the disk aspect ratios ($5/3\simeq\ 1.67$).
Since the gyration frequency of a vortex in a ferromagnetic disk is in a first approximation proportional to the disk aspect ratio \cite{Gus2002}, our data strongly suggests that, as proposed, the
observed signals at opposite current polarities originate from
excitations in the two different disks, the low(high)-frequency mode
corresponding to gyrotropic motion of the vortex in the top (bottom) 
disk. The dominantly excited disk is determined by current polarity in 
agreement with our model.\\
 In summary, we have theoretically and experimentally
demonstrated that windmill modes are quenched in double-vortex
spin-torque nano-oscillators.  The origin is frequency and phase
adaption of the gyrotropic motions in the two disks, which results in a
strong suppression of the gyration radius in one of the disks.
Changing the sign of the exciting current provides an effective mode
selection mechanism, which allows to deliberately choose between
separated frequency bands of the oscillator.
\\
We would like to thank C. Fowley for helpful discussions. A. M. D. acknowledges financial support from the EU project STraDy (MOIF-CT-2006-039772) and Swiss National Science Foundation Ambizione grant PZ00P2\_131808.

\end{document}